\documentclass[11pt,preprint]{aastex}
\usepackage{rotating}

\newcommand{\pivec}{\mbox{\boldmath $\pi$}}

\begin{document}

\title{OGLE-2009-BLG-092/MOA-2009-BLG-137: A Dramatic Repeating Event
With the Second Perturbation Predicted by Real-Time Analysis}

\author{
Y.-H. Ryu\altaffilmark{1},       
C. Han\altaffilmark{1,57,61},         
K.-H. Hwang\altaffilmark{1},    
R. Street\altaffilmark{30,59},       
A. Udalski\altaffilmark{2,56},    
T. Sumi\altaffilmark{3,55},        
A. Fukui\altaffilmark{3,55},   
J.-P. Beaulieu\altaffilmark{4,58}, 
A. Gould\altaffilmark{5,57},    
M. Dominik\altaffilmark{6,59,60}\\  
and\\
F. Abe\altaffilmark{3},        
D.P. Bennett\altaffilmark{7},  
I.A. Bond\altaffilmark{12},
C.S. Botzler\altaffilmark{8},   
K. Furusawa\altaffilmark{3},   
F. Hayashi\altaffilmark{3},    
J.B. Hearnshaw\altaffilmark{9}, 
S. Hosaka\altaffilmark{3},     
Y. Itow\altaffilmark{3},        
K. Kamiya\altaffilmark{3},      
P.M. Kilmartin\altaffilmark{10}, 
A. Korpela\altaffilmark{11},     
W. Lin\altaffilmark{12},         
C.H. Ling\altaffilmark{12},     
S. Makita\altaffilmark{3},      
K. Masuda\altaffilmark{3},      
Y. Matsubara\altaffilmark{3},    
N. Miyake\altaffilmark{3},
Y. Muraki\altaffilmark{13},     
K. Nishimoto\altaffilmark{3},   
K. Ohnishi\altaffilmark{14},    
Y.C. Perrott\altaffilmark{8},   
N. Rattenbury\altaffilmark{8},   
To. Saito\altaffilmark{15},     
%T. Sako\altaffilmark{3},        
L. Skuljan\altaffilmark{12},     
D.J. Sullivan\altaffilmark{11},  
D. Suzuki\altaffilmark{3},      
W.L. Sweatman\altaffilmark{12}, 
P.J. Tristram\altaffilmark{10},  
K. Wada\altaffilmark{13},        
P.C.M. Yock\altaffilmark{8}\\   
(The MOA Collaboration),\\
M.K. Szyma\'nski\altaffilmark{2},  
M. Kubiak\altaffilmark{2},            
G. Pietrzy\'nski\altaffilmark{2,16}, 
I. Soszy\'nski\altaffilmark{2},      
O. Szewczyk\altaffilmark{16},       
{\L}. Wyrzykowski\altaffilmark{17},   
K.\ Ulaczyk\altaffilmark{2}\\        
(The OGLE Collaboration)\\
M. Bos\altaffilmark{51},
G.W. Christie\altaffilmark{18},   
D.L. Depoy\altaffilmark{19},      
A. Gal-Yam\altaffilmark{53},
B.S. Gaudi\altaffilmark{5},      
S. Kaspi\altaffilmark{23},
C.-U. Lee\altaffilmark{20},      
D. Maoz\altaffilmark{23},
J. McCormick\altaffilmark{21},    
B. Monard\altaffilmark{52},
D. Moorhouse\altaffilmark{22},    
R.W.\ Pogge\altaffilmark{5},      
D. Polishook\altaffilmark{23},    
Y. Shvartzvald\altaffilmark{23},
A. Shporer\altaffilmark{23},    
G. Thornley\altaffilmark{22},
J.C. Yee\altaffilmark{5}\\
(The $\mu$FUN Collaboration),\\
M.D. Albrow\altaffilmark{9},
V. Batista\altaffilmark{4,24},    
S. Brillant\altaffilmark{25},     
A. Cassan\altaffilmark{4},       
A. Cole\altaffilmark{27},         
E. Corrales\altaffilmark{4},      
Ch. Coutures\altaffilmark{4},     
S. Dieters\altaffilmark{4,28},    
P. Fouqu\'e\altaffilmark{28},     
J. Greenhill\altaffilmark{27},     
J. Menzies\altaffilmark{29}\\
(The PLANET Collaboration)\\
A. Allan\altaffilmark{48},
D.M. Bramich\altaffilmark{49},
P. Browne\altaffilmark{6,60}, 
K. Horne\altaffilmark{6,58},          
N. Kains\altaffilmark{49,58}
C. Snodgrass\altaffilmark{46,25,60},  
I. Steele\altaffilmark{50},  
Y. Tsapras\altaffilmark{30,53,58}\\
(The RoboNet Collaboration)\\
and\\
V. Bozza\altaffilmark{31,32,33},                 
M.J. Burgdorf\altaffilmark{34,35}, 
S. Calchi Novati\altaffilmark{31,32,33}, 
S. Dreizler\altaffilmark{36}, 
F. Finet\altaffilmark{37}, 
M. Glitrup\altaffilmark{38}, 
F. Grundahl\altaffilmark{38}, 
K. Harps{\o}e\altaffilmark{39}, 
F.V. Hessman\altaffilmark{36}, 
T.C. Hinse\altaffilmark{39,40}, 
M. Hundertmark\altaffilmark{36}
U.G. J{\o}rgensen\altaffilmark{39,41}, 
C. Liebig\altaffilmark{26,6}, 
G. Maier\altaffilmark{26}, 
L. Mancini\altaffilmark{31,32,33,43}, 
M. Mathiasen\altaffilmark{39}, 
S. Rahvar\altaffilmark{44,45}, 
D. Ricci\altaffilmark{37}, 
G. Scarpetta\altaffilmark{31,32,33}, 
J. Skottfelt\altaffilmark{39}, 
J. Surdej\altaffilmark{37}, 
J. Southworth\altaffilmark{47}, 
J. Wambsganss\altaffilmark{40}, 
F. Zimmer\altaffilmark{40}\\
(The MiNDSTEp Collaboration)\\
}

\altaffiltext{1}{Department of Physics, Institute for Astrophysics, Chungbuk National University, Cheongju 371-763, Korea}
\altaffiltext{2}{Warsaw University Observatory, Al. Ujazdowskie 4, 00-478 Warszawa, Poland}
\altaffiltext{3}{Solar-Terrestrial Environment Laboratory, Nagoya University, Nagoya, 464-8601, Japan}
\altaffiltext{4}{Institut d'Astrophysique de Paris, UMR7095 CNRS--Universit\'e Pierre \& Marie Curie, 98 bis boulevard Arago, 75014 Paris, France} 
\altaffiltext{5}{Department of Astronomy, Ohio State University, 140 W. 18th Ave., Columbus, OH 43210, USA}
\altaffiltext{6}{School of Physics \& Astronomy, SUPA, University of St. Andrews, North Haugh, St. Andrews, KY16 9SS, UK}
\altaffiltext{7}{Department of Physics, University of Notre Damey, Notre Dame, IN 46556, USA}
\altaffiltext{8}{Department of Physics, University of Auckland, Private Bag 92019, Auckland, New Zealand}
\altaffiltext{9}{University of Canterbury, Department of Physics and Astronomy, Private Bag 4800, Christchurch 8020, New Zealand}  
\altaffiltext{10}{Mt. John Observatory, P.O. Box 56, Lake Tekapo 8770, New Zealand} 
\altaffiltext{11}{School of Chemical and Physical Sciences, Victoria University, Wellington, New Zealand} 
\altaffiltext{12}{Institute of Information and Mathematical Sciences, Massey University, Private Bag 102-904, North Shore Mail Centre, Auckland, New Zealand} 
\altaffiltext{13}{Department of Physics, Konan University, Nishiokamoto 8-9-1, Kobe 658-8501, Japan} 
\altaffiltext{14}{Nagano National College of Technology, Nagano 381-8550, Japan} 
\altaffiltext{15}{Tokyo Metropolitan College of Industrial Technology, Tokyo 116-8523, Japan} 
\altaffiltext{16}{Universidad de Concepci\'on, Departamento de Fisica, Casilla 160-C, Concepci\'on, Chile} 
\altaffiltext{17}{Institute of Astronomy Cambridge University, Madingley Road, CB3 0HA Cambridge, UK} 
\altaffiltext{18}{Auckland Observatory, Auckland, New Zealand} 
\altaffiltext{19}{Department of Physics, Texas A\&M University, College Station, TX, USA} 
\altaffiltext{20}{Korea Astronomy and Space Science Institute, Daejeon 305-348, Korea} 
\altaffiltext{21}{Farm Cove Observatory, Pakuranga, Auckland} 
\altaffiltext{22}{Kumeu Observatory, Kumeu, New Zealand} 
\altaffiltext{23}{School of Physics and Astronomy, Tel-Aviv University, Tel Aviv 69978, Israel} 
\altaffiltext{24}{Institut d\'Astrophysique de Paris, UPMC Univ Paris 06, UMR7095, F-75014, Paris, France} 
\altaffiltext{25}{European Southern Observatory, Casilla 19001, Vitacura 19, Santiago, Chile} 
\altaffiltext{26}{Astronomisches Rechen-Institut, Zentrum f\"ur Astronomie der Universit¡§at Heidelberg, M\"onchhofstrasse 12-14, 69120 Heidelberg, Germany} 
\altaffiltext{27}{School of Math and Physics, University of Tasmania, Private Bag 37, GPO Hobart, Tasmania 7001, Australia} 
\altaffiltext{28}{LATT, Universit\'e de Toulouse, CNRS, 14 Avenue Edouard Belin, 31400 Toulouse, France} 
\altaffiltext{29}{South African Astronomical Observatory, P.O. Box 9 Observatory 7935, South Africa} 
\altaffiltext{30}{Las Cumbres Observatory Global Telescope Network, 6740B Cortona Dr, Suite 102, Goleta, CA 93117, USA} 

\altaffiltext{31}{Universit\`{a} degli Studi di Salerno, Dipartimento di Fisica "E.R.~ Caianiello", Via Ponte Don Melillo, 84085 Fisciano (SA), Italy}
\altaffiltext{32}{INFN, Gruppo Collegato di Salerno, Sezione di Napoli, Italy}
\altaffiltext{33}{Istituto Internazionale per gli Alti Studi Scientifici (IIASS), Via G.\ Pellegrino 19, 84019 Vietri sul Mare (SA), Italy}
\altaffiltext{34}{Deutsches SOFIA Institut, Universit\"{a}t Stuttgart, Pfaffenwaldring 31, 70569 Stuttgart, Germany}
\altaffiltext{35}{SOFIA Science Center, NASA Ames Research Center, Mail Stop N211-3, Moffett Field CA 94035, United States of America}
\altaffiltext{36}{Institut f\"{u}r Astrophysik, Georg-August-Universit\"{a}t, Friedrich-Hund-Platz 1, 37077 G\"{o}ttingen, Germany}
\altaffiltext{37}{Institut d'Astrophysique et de G\'{e}ophysique, All\'{e}e du 6 Ao\^{u}t 17, Sart Tilman, B\^{a}t.\ B5c, 4000 Li\`{e}ge, Belgium}
\altaffiltext{38}{Department of Physics \& Astronomy, Aarhus University, Ny Munkegade 120, 8000 {\AA}rhus C, Denmark}
\altaffiltext{39}{Niels Bohr Institutet, K{\o}benhavns Universitet, Juliane Maries Vej 30, 2100 K{\o}benhavn {\O}, Denmark}
\altaffiltext{40}{Armagh Observatory, College Hill, Armagh, BT61 9DG, UK}
\altaffiltext{41}{Centre for Star and Planet Formation, K{\o}benhavns Universitet, {\O}ster Voldgade 5-7, 1350 K{\o}benhavn {\O}, Denmark}
\altaffiltext{42}{Astronomisches Rechen-Institut, Zentrum f\"{u}r Astronomie der Universit\"{a}t Heidelberg (ZAH),  M\"{o}nchhofstr.\ 12-14, 69120 Heidelberg, Germany}
\altaffiltext{43}{Dipartimento di Ingegneria, Universit\`{a} del Sannio, Corso Garibaldi 107, 82100 Benevento, Italy}
\altaffiltext{44}{Department of Physics, Sharif University of Technology, P.O.~Box 11365--9161, Tehran, Iran}
\altaffiltext{45}{School of Astronomy, IPM (Institute for Studies in Theoretical Physics and Mathematics), P.O. Box 19395-5531, Tehran, Iran}
\altaffiltext{46}{Max Planck Institute for Solar System Research, Max-Planck-Str. 2, 37191 Katlenburg-Lindau, Germany}
\altaffiltext{47}{Astrophysics Group, Keele University, Staffordshire, ST5 5BG, UK}
\altaffiltext{48}{School of Physics, University of Exeter, Stocker Road, Exeter, Devon, EX4 4QL, UK}
\altaffiltext{49}{European Southern Observatory, Karl-Schwarzschild-Stra{\ss}e 2, 85748 Garching bei M{\"u}nchen, Germany}
\altaffiltext{50}{Astrophysics Research Institute, Liverpool John Moores University, Egerton Wharf, Birkenhead CH41 1LD, UK}
\altaffiltext{51}{Molehill Astronomical Observatory, North Shore, New Zealand}
\altaffiltext{52}{Bronberg Observatory, Pretoria, South Africa} 
\altaffiltext{53}{School of Mathematical Sciences, Queen Mary, University of London, London E1 4NS}
\altaffiltext{53}{Benoziyo Center for Astrophysics, the Weizmann Institute, Israel} 
\altaffiltext{55}{The MOA Collaboration}
\altaffiltext{56}{The OGLE Collaboration}
\altaffiltext{57}{The $\mu$FUN Collaboration}
\altaffiltext{58}{The PLANET Collaboration}
\altaffiltext{59}{The RoboNet Collaboration}
\altaffiltext{60}{MiNDSTEp Collaboration}
\altaffiltext{61}{Corresponding author}

\begin{abstract}
We report the result of the analysis of a dramatic repeating 
gravitational microlensing event OGLE-2009-BLG-092/MOA-2009-BLG-137, 
for which the light curve is characterized by two distinct peaks 
with perturbations near both peaks.  We find that the event 
is produced by the passage of the source trajectory over the 
central perturbation regions associated with the individual 
components of a wide-separation binary.  The event is special 
in the sense that the second perturbation, occurring $\sim 100$ 
days after the first, was predicted by the real-time analysis 
conducted after the first peak, demonstrating that real-time 
modeling can be routinely done for binary and planetary events.  
With the data obtained from follow-up observations covering the 
second peak, we are able to uniquely determine the physical 
parameters of the lens system.  We find that the event occurred 
on a bulge clump giant and it was produced by a binary lens 
composed of a K and M-type main-sequence stars.  The estimated 
masses of the binary components are $M_1=0.69 \pm 0.11\ M_\odot$ 
and $M_2=0.36\pm 0.06\ M_\odot$, respectively, and they are 
separated in projection by $r_\perp=10.9\pm 1.3\ {\rm AU}$.  
The measured distance to the lens is $D_{\rm L}=5.6 \pm 0.7\ 
{\rm kpc}$.  We also detect the orbital motion of the lens system.
\end{abstract}

\keywords{gravitational lensing}

\section{Introduction}

When a foreground astronomical object is closely aligned to a 
background star, the light from the background star (source) is 
amplified by the gravity of the foreground object (lens).  The 
magnification of this gravitational lensing phenomenon depends 
on the projected separation between the lens and source star.  
With the change of the separation, the lensing magnification 
varies in time.  For an event caused by a single-mass lens, the 
brightness variation is characterized by its non-repeating 
symmetric light curve \citep{paczynski86}.

When a star is gravitationally magnified by a lens composed of 
two masses (binary lens), the resulting light curves become 
complicated due to the non-linear nature of binary-lensing 
magnifications \citep{schneider86}.  The most important feature 
of binary lensing is caustics, which represent the set of source 
positions at which the lensing magnification of a point source 
becomes infinite.  Caustics form a single or multiple sets of 
closed curves each of which is composed of concave curves that 
meet at cusps.  The number, shape, size, and locations of 
caustic curves vary depending on the separation and mass ratio 
between the binary components.  As a result, light curves of 
binary-lensing events exhibit great diversity \citep{erdl93, 
mao91}.

In current microlensing experiments, events are observed from 
the combination of survey and follow-up observations.  Survey 
observations,  e.g., OGLE \citep{udalski05} and MOA \citep{sumi10}, 
are operated in order to maximize the event rate by monitoring 
a large area of sky toward the Galactic bulge on a roughly 
nightly basis using large-format cameras.  On the other hand, 
follow-up observations, e.g., $\mu$FUN \citep{gould06},
PLANET \citep{beaulieu06}, and RoboNet \citep{tsapras09}, are 
focused on events alerted by survey observations to densely 
cover various anomalies including planet-induced perturbations. 
However, the limited number of telescopes available for follow-up 
observations restricts the number of events that can be monitored 
at any given time.  For the efficient use of telescopes for 
follow-up observations, then, it is important to judge which 
events should be focused upon among the events alerted by survey 
observations. This judgment can be done based on real-time modeling 
of events.  Real-time modeling is also important to judge the time 
and duration of follow-up observations.  Extended coverage of events 
is often needed to determine the physical parameters of lenses by 
measuring subtle deviations caused by long-lasting effects.  In 
addition, planets and binaries often induce multiple perturbations, 
and resolving additional perturbations is very critical for accurate 
and precise characterization of lenses.  Real-time modeling helps 
to judge how long and when intensive follow-up observations should 
be carried out.  Despite its importance, routine real-time modeling 
of binary and planetary lensing events has been difficult due to 
the large number of parameters to be included in modeling combined 
with the complexity of $\chi^2$ surface of the parameter space.  
With the development of efficient codes from the efforts of 
theoretical studies on binary-lensing phenomenology, however, it 
is now possible to routinely release models of light curves just 
after or even during the progress of perturbations.\footnote{Since 
2009, real-time models of most of anomalous events alerted by 
survey observations are posted on the webpage 
{\tt http://astroph.cbnu.ac.kr/$\sim$cheongho/modelling/model\_"year".html},
where ``year'' corresponds to the year that events were discoverd.
}

In this paper, we present results of the analysis of a dramatic 
repeating event for which the light curve is characterized by 
two distinct peaks with perturbations near both peaks. 
The event is highlighted because the second perturbation, 
which occurred $\sim 100$ days after the first peak, was 
predicted by the real-time analysis conducted after the 
first perturbation.  With the dense coverage of the second 
peak by follow-up observations based on the real-time modeling, 
we are able to measure the physical parameters of the lens.

% Figure 1 ----------------------------------------------------
\begin{figure*}[th]
\epsscale{0.8}
\plotone{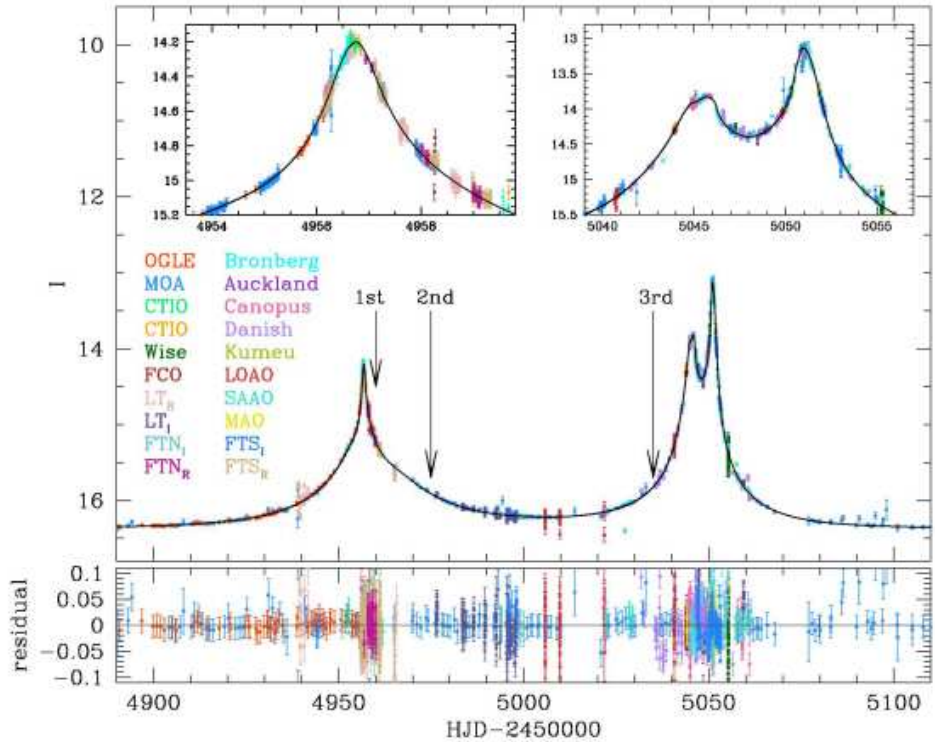}
\caption{\label{fig:one}
Light curve of the microlensing event OGLE-2009-BLG-092/MOA-2009-BLG-137.  
We note that the MOA data except during the perturbation regions are binned 
(by 1 day) for clarity.  The insets in the upper panel show enlargements 
of the individual perturbation regions.  The lower panel shows the 
residual from the best-fit model.  The arrows represent the times 
when the real-time analyses were conducted.
}\end{figure*}
% -------------------------------------------------------------

\section{Observations and Real-Time Modeling}

The event OGLE-2009-BLG-092/MOA-2009-BLG-137 occurred on a Galactic 
bulge star located at $(\alpha,\delta)_{2000}=(17^{\rm h}51^{\rm m}
37^{\rm s} \hskip-2pt.95, -29^\circ 32' 43.23'')$, which corresponds 
to the Galactic coordinates $(l,b)=(0.15^\circ,-1.44^\circ)$.  The 
event was independently detected by the Optical Gravitational Lensing 
Experiment (OGLE) and Microlensing Observations in Astrophysics (MOA) 
groups using the 1.3 m Warsaw telescope of Las Campanas Observatory 
in Chile and 1.8 m telescope of Mt.\ John Observatory in New Zealand, 
respectively.

%The event was alerted as a high-magnification event, which has 
%a high chance to detect planet-induced anomalies \citep{griest98}.  
%Based on the alert, the peak of the light curve 

An anomaly alert was issued on on 2009 May 4.
Based on the alert, the event was intensively 
observed by follow-up groups of the Microlensing Follow-Up Network 
($\mu$FUN), the Probing Lensing Anomalies Network (PLANET), and the 
RoboNet collaborations.  The telescopes used for these follow-up 
observations include the $\mu$FUN 1.3 m SMARTS telescope of CTIO 
in Chile, $\mu$FUN 0.4 m of Auckland Observatory, $\mu$FUN 0.4 m 
of Farm Cove Observatory (FCO) in New Zealand, $\mu$FUN 1.0 m Mt. 
Lemmon Observatory (LOAO) in Arizona, USA, $\mu$FUN 0.4 m of 
Bronberg Observatory in South Africa, RoboNet 2.0 m Liverpool 
Telescope (LT) in La Palma, Canary Islands, RoboNet 2.0 m Faulkes 
North (FTN) in Hawaii, RoboNet 2.0 m Faulkes South (FTS) in Australia, 
and PLANET 1.0 m of Mt.\ Canopus Observatory in Australia.  Dense 
coverage by the survey and follow-up observations revealed an 
anomaly near the peak of the light curve.

Just after the perturbation, real-time analysis of the event was
conducted using data available at that moment.  Modeling showed 
that the perturbations were produced by a binary lens.  The 
projected separation of the binary system was degenerate between 
two values, one larger than the Einstein radius and the other 
smaller.  This degeneracy is known as the close/wide binary degeneracy 
\citep{dominik99b}.  It was also found that if the perturbation was 
produced by a wide binary, the source trajectory would pass close 
to the caustic associated with the other component of the binary, 
and thus there would be an additional perturbation approximately 
$\sim 100$ days after the first perturbation.

Another modeling conducted at around ${\rm HJD}=2454975$ with 
additionally acquired data resolved the close/wide binary degeneracy 
and finally predicted the second anomaly.  The region between the 
two peaks was expected to vary smoothly without any major perturbation.
In addition, the time gap between the two peaks is too long to be 
continuously monitored by follow-up observations.  Therefore, 
observations during this period were conducted mostly by survey 
groups.

A third modeling was conducted at ${\rm HJD}\sim 2455035$ to 
precisely predict the time of the second perturbation.  From this, 
the second perturbation was predicted to occur at HJD$\sim$2455045.
With this prediction, another follow-up campaign was prepared to 
cover the second perturbation.  In addition to the telescopes used 
for the follow-up observations of the first perturbation, additional 
telescopes participated in the observations of the second peak. 
These include the PLANET 1.54 m Danish Telescope of La Silla 
Observatory of the Microlensing Network for the Detections of Small 
Terrestrial Exoplanets (MiNDSTEp) group, $\mu$FUN 1.0 m of Wise 
Observatory in Israel, $\mu$FUN 0.36 m of Kumeu Observatory, 
$\mu$FUN 0.3 m of Molehill Astronomy Observatory (MAO) in New 
Zealand, and PLANET 1.0 m of SAAO in South Africa.  From this
campaign, the second peak was also densely resolved.

Figure \ref{fig:one} shows the light curve of 
OGLE-2009-BLG-092/MOA-2009-BLG-137.  In the light curve, the MOA 
data except during the perturbation regions are binned by 1 day for 
clarity but the modeling is based on unbinned data.  The insets show 
enlargements of the perturbation regions around the individual peaks.  
We mark the times when the real-time analyses were conducted.

\headsep=30pt
\begin{sidewaystable}[ht]
\caption{Fit Parameters\label{table:one}}
\begin{tabular}{lccccc}
\hline\hline 
\multicolumn{1}{c}{} &
\multicolumn{5}{c}{model} \\
\multicolumn{1}{c}{parameter} &
\multicolumn{1}{c}{standard} &
\multicolumn{2}{c}{parallax} &
\multicolumn{2}{c}{parallax+orbital motion} \\
\multicolumn{1}{c}{} &
\multicolumn{1}{c}{} &
\multicolumn{1}{c}{$u_0>0$} &
\multicolumn{1}{c}{$u_0<0$} &
\multicolumn{1}{c}{$u_0>0$} &
\multicolumn{1}{c}{$u_0<0$} \\
\hline
% ----------------------------------------
$\chi2/{\rm dof}$              & 6639.03/6569        & 6616.06/6567        & 6598.27/6567       & 6510.61/6565        & {\bf 6503.02/6565}        \\
$t_0$ (HJD)                    & 4990.441$\pm$ 0.053 & 4990.501$\pm$ 0.047 & 4990.106$\pm$0.054 & 4988.473$\pm$ 0.058 & {\bf 4987.744$\pm$ 0.047} \\
$u_0$                          & -0.062$\pm$ 0.001   & 0.060$\pm$ 0.001    &-0.063$\pm$0.001    & 0.053$\pm$ 0.001    & {\bf -0.051$\pm$ 0.001}   \\
$t_{\rm E}$ (days)             & 35.53$\pm$ 0.01     & 35.44$\pm$ 0.04     & 35.48$\pm$0.03     & 35.23$\pm$ 0.05     & {\bf 35.25$\pm$ 0.03}     \\
$s$                            & 2.921$\pm$ 0.001    & 2.923$\pm$ 0.001    & 2.927$\pm$0.001    & 2.909$\pm$ 0.002    & {\bf 2.8946$\pm$ 0.001}    \\
$q$                            & 0.562$\pm$ 0.001    & 0.565$\pm$ 0.001    & 0.557$\pm$0.002    & 0.530$\pm$ 0.001    & {\bf 0.518$\pm$ 0.001}    \\
$\alpha$ (rad)                 & 6.2466$\pm$ 0.0001  & 0.0343$\pm$ 0.0003  & 6.2410$\pm$0.0002  & 0.0411$\pm$ 0.0003  & {\bf 6.2420$\pm$ 0.0002}  \\
$\rho_\star$                   & 0.0101$\pm$ 0.0001  & 0.0100$\pm$ 0.0001  & 0.0094$\pm$0.0001  & 0.0091$\pm$ 0.0001  & {\bf 0.0090$\pm$ 0.0001}  \\
$\pi_{{\rm E},N}$              & --                  & -0.055$\pm$ 0.009   &-0.125$\pm$0.005    & -0.018$\pm$ 0.016   & {\bf 0.059$\pm$ 0.008}    \\
$\pi_{{\rm E},E}$              & --                  & -0.019$\pm$ 0.013   & 0.057$\pm$0.009    & -0.039$\pm$ 0.013   & {\bf -0.054$\pm$ 0.012}   \\
$\dot{s}$ ($t_{\rm E}^{-1}$)   & --                  & --                  & --                 & 0.0315$\pm$ 0.0016  & {\bf 0.0464$\pm$ 0.0013}  \\
$\omega$ ($t_{\rm E}^{-1}$)    & --                  & --                  & --                 & -0.0068$\pm$ 0.0007 & {\bf 0.0099$\pm$ 0.0006} \\
% -------------------------------------
\hline
\end{tabular}
\\The parameters of the best-fit solution are marked in bold fonts.
\end{sidewaystable}

\section{Modeling}

Due to their great diversity, describing light curves of binary-lensing 
events requires to include many parameters.  The basic structure of 
light curves of binary-lens events is characterized by six lensing 
parameters.  The first set of three parameters are needed to describe 
light curves of standard single-lens events:  the time required for 
the source to transit the Einstein radius, $t_{\rm E}$ (Einstein time 
scale), the time of the closest lens-source approach, $t_0$, and the 
lens-source separation in units of $\theta_{\rm E}$ at that time, 
$u_0$ (impact parameter).  Describing the deviation caused by the 
lens binarity requires an additional set of three binary-lensing 
parameters: the mass ratio between the lens components, $q$, the 
projected binary separation in units of the Einstein radius, $s$, 
and the angle of the source trajectory with respect to the binary 
axis, $\alpha$.

In addition to these basic parameters, additional parameters are 
needed to describe detailed structures of lensing light curves.
The event OGLE-2009-BLG-092/MOA-2009-BLG-137 exhibits caustic-induced 
perturbations at both peaks and thus it is required to consider the 
finite-source effect to describe deviations occurring when the 
source approaches and crosses over caustics \citep{nemiroff94, 
witt94, gould94}.  The finite-source effect is parameterized by 
the ratio of the source radius $\theta_\star$ to the Einstein 
radius $\theta_{\rm E}$, i.e.\ $\rho_\star=\theta_{\star}/
\theta_{\rm E}$ (normalized source radius).

Due to the large time gap between the two peaks of the light curve, 
the relative lens-source motion may deviate from a rectilinear one due 
to the  acceleration of the observer's motion induced by the Earth's 
orbital motion around the Sun \citep{refsdal66, gould92, smith03}.
We consider this so-called ``parallax effect'' in the modeling by 
including the two parallax parameters $\pi_{{\rm E},N}$ and 
$\pi_{{\rm E},E}$, which are the two components of the microlensing
parallax vector $\pivec_{\rm E}$ projected on the sky in the direction 
of north and east celestial coordinates.  The direction of this vector 
is that of the lens-source relative motion in the frame of the Earth 
at $t_0$.\footnote{We set ``$t_0$'' as the time of the source star's 
closest approach to the center of mass of the binary lens.}

We also check the possibility of the change of lens positions 
caused by its orbital motion.  The orbital motion has two effects 
on lensing magnifications.  One causes the binary axis to rotate 
or, equivalently, makes the source trajectory angle, $\alpha$, 
change in time.  The other effect is causing the separation between 
the binary components to vary in time \citep{dominik98, ioka99, 
albrow00}.  The change in the binary separation alters the shape 
of the caustic in the course of the event.  To the first order, 
the orbital effect is parameterized by 
\begin{equation}
\alpha(t)=\alpha(t_0)+\omega\left( {t-t_0\over t_{\rm E}}\right)
\label{eq1}
\end{equation}
and 
\begin{equation}
s(t)=s(t_0)+\dot{s}\left( {t-t_0\over t_{\rm E}}\right),
\label{eq2}
\end{equation}
where the orbital lensing parameters $\omega$ and $\dot{s}$ represent 
the rates of change in the source trajectory angle and the projected 
binary separation, respectively.

Due to the sheer size of the parameter space, it is difficult to 
find binary-lensing solutions from brute-force searches.  Searches 
for solutions becomes further hampered by the complexity of the 
$\chi^2$ surface.  This complexity implies that even if a solution 
that seemingly describes an observed light curve is found, it is 
difficult to be sure that all possible minima have been investigated 
\citep{dominik99a, dominik99b} and thus a simple downhill approach 
cannot be used.  To avoid these difficulties, we use a hybrid approach
in whcih grid searches are conducted over the space of a subset of
parameters and the remaining parameters are allowed to vary so that 
the model light curve results in minimum $\chi^2$ at each grid point.
See also \citet{dong06} and \citet{bennett10a}.
We choose $s$, $q$, and $\alpha$ as grid parameters because they are 
related to the features of light curves in a complicated pattern such 
that a small change in these parameters can result in dramatic changes 
in the resulting light curves. On the other hand, the other parameters 
are more directly related to the observed light curve features. We use 
a Markov Chain Monte Carlo method for $\chi^2$ minimization.  Once the 
solutions of the individual grid points are determined, the best-fit 
model is obtained by comparing the $\chi^2$ minima of the individual 
grid points.

In addition to the difficulties mentioned above, binary-lens
modeling suffers from an additional difficulty that arises due to 
large computations required for modeling.  Most binary-lensing 
events exhibit perturbations induced by caustic crossings 
or approaches during which the finite-source effect is important.  
Calculating finite-source magnifications requires a numerical method 
for which heavy computations are needed.  Considering that modeling 
requires to produce many light curves of trial models, it is important 
to apply an efficient method for magnification calculations.  In our 
modeling, we use a customized version of the inverse ray-shooting 
method to calculate finite-source magnifications. In the usual 
ray-shooting method, a large number of rays are uniformly shot from 
the image plane, bent according to the lens equation, and land on 
the source plane. Then, the lensing magnification corresponding to 
the location of a finite source is computed by comparing the number 
density of rays on the source surface with the density on the image 
plane. The main shortcoming of this method is that only a small 
fraction of rays land on the source surface and most of the rest 
of the rays are not used for magnification computations.  We reduce 
computation time by minimizing wasted rays.  For this, we first 
make grids on the image plane.  We then find the image positions 
corresponding to the individual positions of the envelope of the 
source star and then register the grids corresponding to the image 
positions.  We minimize the number of rays by restricting the region 
of ray shooting only to the registered grids on the image plane.  
This scheme is similar to that of \citet{rattenbury02}.
We set the width of grids slightly bigger than the source star. If 
the width is too small, the region inside the image may not be 
registered.  If the width is too big, the fraction of rays not 
arriving on the source surface will increase. We find that the 
optimal grid width corresponds to the diameter of the source. 
This is because lensing-induced distortions always result in 
images slimmer than the source and thus grids with a width of 
the source diameter fill the images, thereby minimizing wasted 
rays.  To further speed up the computation, we use the finite-source 
magnification calculations based on the numerical ray-shooting 
method only in the region near the caustic, and a simple semi-analytic 
hexadecapole approximation \citep{pejcha09, gould08} is used in 
other part of light curves.

% Figure 2 ----------------------------------------------------
\begin{figure*}[t]
\epsscale{0.7}
\plotone{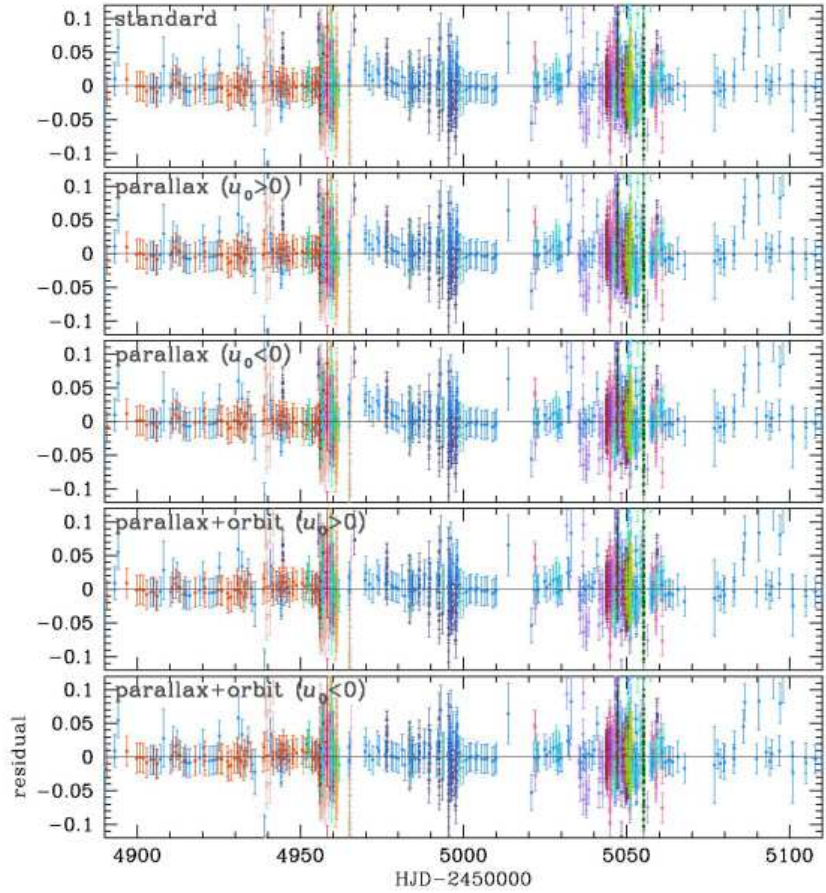}
\caption{\label{fig:two}
Residuals of data from various models.
}\end{figure*}
% -------------------------------------------------------------

% Figure 3 ----------------------------------------------------
\begin{figure*}[t]
\epsscale{0.7}
\plotone{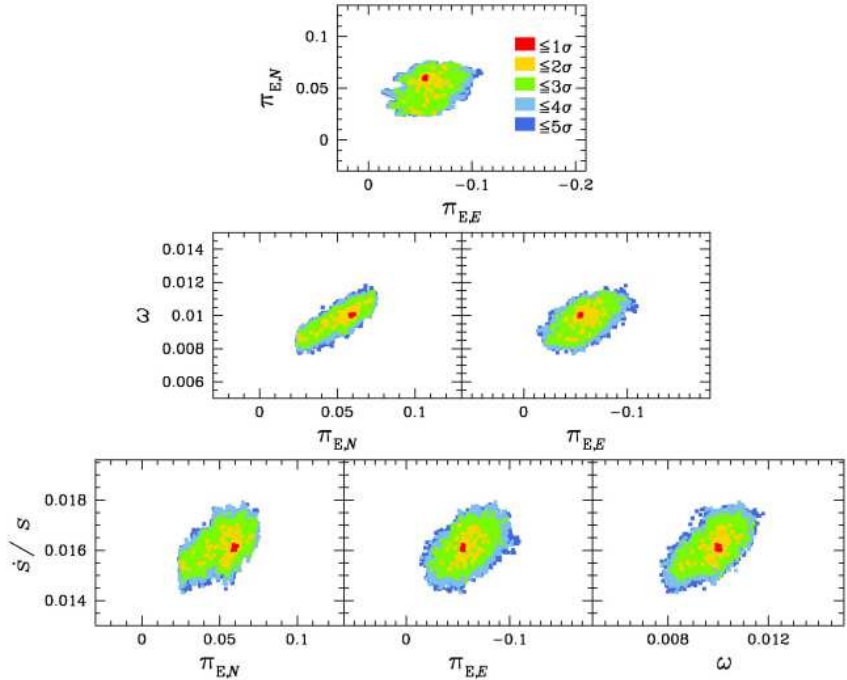}
\caption{\label{fig:three}
Contours of $\chi^2$ in the space of parallax and orbital lensing 
parameters.
}\end{figure*}
% -------------------------------------------------------------

% Figure 4 ----------------------------------------------------
\begin{figure*}[th]
\epsscale{0.8}
\plotone{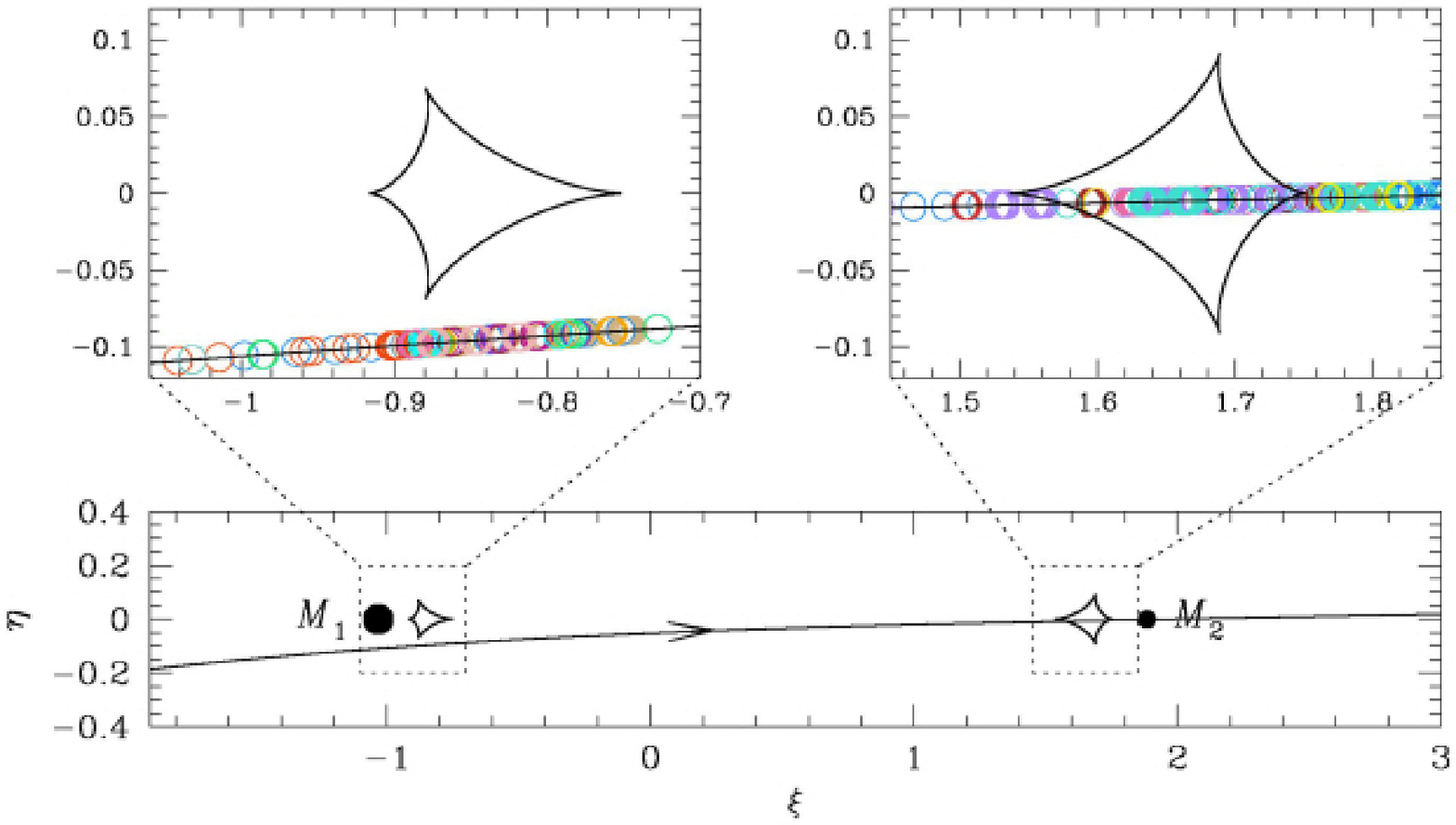}
\caption{\label{fig:four}
Geometry of the lens system responsible for the microlensing event 
OGLE-2009-BLG-092/MOA-2009-BLG-137.  The filled circles represent 
the locations of the lens components, where $M_1$ is the more massive 
component.  The two closed curves are the caustics formed by the 
binary lens.  The curve with an arrow represents the source trajectory.  
The two upper panels show enlargements of the region around the 
individual caustics, which correspond to the times of the peaks in 
the light curve at HJD$\sim$2454959 and 2455048, respectively.
The open circles on the source trajectory 
represent the source star at the times of observations, where the 
size indicates its finite size and the colors correspond to those 
of the data points of different observatories used in Fig.\ \ref{fig:one}.  
All lengths are normalized by the Einstein radius corresponding to the 
total mass of the binary.
}\end{figure*}
% -------------------------------------------------------------

\section{Results}

In Table \ref{table:one}, we summarize the results of modeling.  
We test 5 different models.  The first model is based on a static 
binary lens (no orbital effect) without the parallax effect (standard 
model).  The second model includes the parallax effect.  Finally, 
the orbital motion of the lens is additionally considered in the 
third model.  When the parallax or orbital motion is considered, a 
pair of solutions resulting from the mirror-image source trajectories 
with impact parameters $u_0>0$ and $u_0<0$ result in slightly different 
light curves due to the asymmetry of the source trajectories with 
respect to the binary axis.  We, therefore, check both models with 
$u_0>0$ and $u_0<0$ whenever the parallax or orbital effect is 
considered.

From the table, it is found that the model including both the parallax 
and orbital effects resulting from a source trajectory with $u_0<0$ 
provides the best fit to the observed light curve.  It is also found 
that the parallax effect improves the fit by $\Delta\chi^2/{\rm dof}
=40.8/6567$, while the orbital effect further improves the fit by 
$\Delta\chi^2/{\rm dof} =136.0/6565$.  The differences in the goodness 
of fit between the solutions can be seen in Figure \ref{fig:two}, where 
we present the residuals of data from the individual models.  One finds 
that the ``parallax + orbit'' solution removes the systematic residuals 
that are present in the other solutions.  We note that the amount of 
parallax $\pi_{\rm E} = (\pi_{{\rm E},E}^2+\pi_{{\rm E},N}^2)^{1/2}=
0.080$ is substantially smaller than the typical values of events for 
which parallaxes are measured.  The orbit-induced changes of the binary 
separation during the time gap $\Delta t$ between the two peaks of 
$\Delta s= \dot{s}(\Delta t/t_{\rm E})\sim 0.13$ and the source trajectory 
angle of $\Delta \alpha= \omega (\Delta t/t_{\rm E})\sim 1.6^\circ$ are 
very small.  Despite their small amplitudes, both parallax and orbital 
effects are measurable thanks to the dense coverage of the second 
perturbation from follow-up observations combined with the fact that 
the event had a long effective time scale and there were two disconnected 
deviations.  Figure \ref{fig:three} presents the contours of $\chi^2$ in 
the space of the combinations of the parallax and orbital lensing parameters.  
We note that the tangential and radial velocities of the companion 
relative to the primary are $v_{\rm t}= r_\perp\omega$ and $v_{\rm r}
=r_\perp(\dot{s}/s)$, respectively, where $r_\perp$ represents the 
projected binary separation.  We, therefore, set the ordinate of the 
lower panels as $\dot{s}/s$ so that it is in the same order of $\omega$.

In Figure \ref{fig:one}, we present the model light curve of the 
best-fit solution.  In Figure \ref{fig:four}, we also present the 
geometry of the lens system based on the best-fit solution.  In 
the figure, the filled circles represent the locations of the lens 
components, where $M_1$ is the heavier component, the two closed 
curves are the caustics, and the curve with an arrow represents 
the source trajectory.  The two upper panels show enlargements of 
the regions around the individual caustics.  The open circles on 
the source trajectory represent the source star at the times of 
observations where the size indicates its finite size and the colors 
correspond to those of the data points of different observatories 
adopted in Figure \ref{fig:one}.  All lengths are normalized by the 
Einstein radius corresponding to the total mass of the binary.

% Figure 5 ----------------------------------------------------
\begin{figure*}[t]
\epsscale{0.6}
\plotone{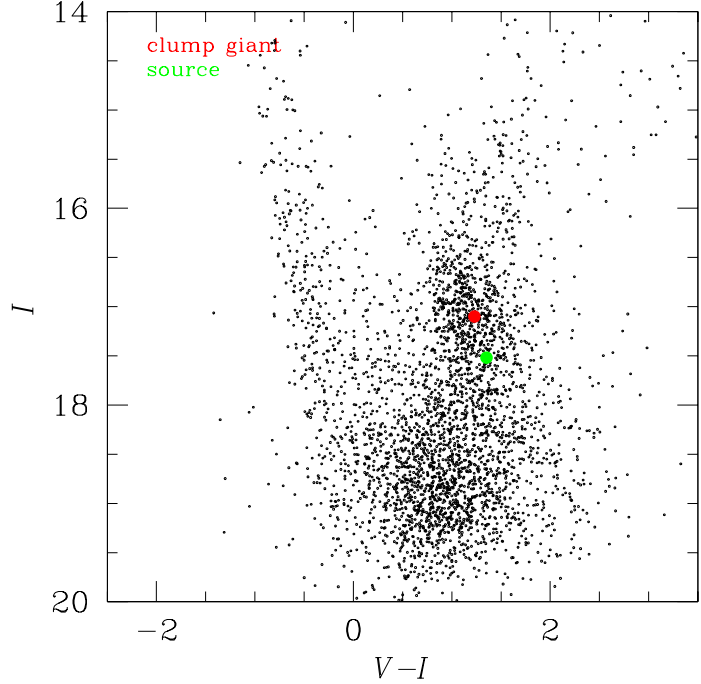}
\caption{\label{fig:five}
Instrumental color-magnitude diagram of stars in the field 
containing the source star of OGLE-2009-BLG-092/MOA-2009-BLG-137.
}\end{figure*}
% -------------------------------------------------------------

To determine the physical parameters of the lens system,  it is 
required to measure both the Einstein radius and the lens parallax.  
The lens parallax is directly measured from modeling.  The Einstein 
radius is measured from the  normalized source radius $\rho_\star$ 
combined with the angular size of the source star, $\theta_\star$. 
The angular source size is estimated based on the de-reddened 
magnitude $I_0$ and color $(V-I)_0$ of the source star measured 
from the offset between the source and the centroid of clump 
giants in the instrumental color-magnitude diagram under the 
assumption that source and clump giants experience the same amount 
of extinction \citep{yoo04}.  In Figure \ref{fig:five}, we present 
the location of the source in the color-magnitude diagram constructed 
by using the $V$ and $I$ band images taken from CTIO.  With the known 
clump centroid of $[(V-I)_0,I_0]_{\rm c}=(1.04,14.27)$ and the measured 
offsets of $\Delta(V-I)=(V-I)_{\rm S}-(V-I)_{\rm c}=0.123$ and 
$\Delta I= I_{\rm S}-I_{\rm c}=0.429$, the de-reddened color and 
magnitude of the source star are measured as $[(V-I)_0,I_0]_{\rm S}
=(1.16,14.69)$, respectively.  Here we adopt the distance to the 
clump of 8.0 kpc toward the field which is estimated by using the 
Galactic model of \citet{han03}.  Then, the angular source size is 
determined by first transforming from $(V-I)_0$ to $(V-K)_0$ using 
the color-color relation of \citet{bessell88} and then applying the 
relation between $(V-K)_0$ and the angular stellar radius of 
\citet{kervella04}.  The determined angular radius of the source 
star is $\theta_\star = 6.11\pm 0.53\ \mu{\rm as}$, implying that 
the source star is a Galactic bulge clump giant star.  The uncertainty 
of $\theta_\star$ is estimated from the combination of those of the 
colors and magnitudes of the source and the clump centroid and an 
additional 7\% intrinsic error in the conversion process from the 
color to the source radius \citep{yee09}.  Then, the Einstein radius 
is measured as
\begin{equation}
\theta_{\rm E}={\theta_\star\over \rho_\star}=0.68\pm0.06\ {\rm mas},
\label{eq3}
\end{equation}
Combining this with the Einstein time scale yields the relative 
proper motion between the lens and source of 
\begin{equation}
\mu = {\theta_{\rm E}\over t_{\rm E}}
=7.01 \pm 0.61 \ {\rm mas}\ {\rm yr}^{-1}.
\label{eq4}
\end{equation}

With the measured Einstein radius and lens parallax, the mass of 
the lens and distance to the lens are determined, respectively, as
\begin{equation}
M={\theta_{\rm E}\over \kappa\pi_{\rm E}},\qquad
D_{\rm L}={{\rm AU}\over \pi_{\rm E}\theta_{\rm E}+\pi_{\rm S}},
\label{eq5}
\end{equation}
where $\kappa=4G/(c^2{\rm AU})$, $\pi_{\rm S}={\rm AU}/D_{\rm S}$ 
is the parallax of the source, $D_{\rm L}$ and $D_{\rm S}$ 
represent the distances to the lens and source, respectively.  For 
the best-fit model, the determined values are
\begin{equation}
M=1.04 \pm 0.16 \ M_\odot
\label{eq6}
\end{equation}
and
\begin{equation}
D_{\rm L}=5.6 \pm 0.7\ {\rm kpc},
\label{eq7}
\end{equation}
respectively.  From the normalized separation together with the 
physical Einstein radius $r_{\rm E}=D_{\rm L} \theta_{\rm E}$, 
the projected separation between the binary companions is 
estimated as 
\begin{equation}
r_\perp=s r_{\rm E}=10.9\pm 1.3\ {\rm AU}.  
\label{eq8}
\end{equation}
With the known mass ratio, the masses of the individual binary 
components are estimated, respectively, as 
\begin{equation}
M_1={M \over 1+q}=0.69 \pm 0.11\ M_\odot 
\label{eq9}
\end{equation}
and 
\begin{equation}
M_2= {q \over 1+q}M=0.36 \pm 0.06\ M_\odot. 
\label{eq10}
\end{equation}
Therefore, the event OGLE-2009-BLG-092/MOA-2009-BLG-137 occurred 
on a bulge clump giant and it was produced by a binary lens 
composed of a K and M-type main-sequence stars.

In principle, it is possible to constrain the physical orbital
parameters such as the semi-major axis, orbital period, and 
inclination of the orbital plane from the determined orbital 
lensing parameters of $\omega$ and $\dot{s}$ \citep{dong09, 
bennett10b}.   However, we find that this is difficult because 
the magnitudes of the changes of the binary separation and 
source trajectory angle are too small.  Nevertheless, it is 
still possible to check the consistency of the orbital parameters 
using the parameter 
\begin{equation}
\eta=
{(r_\perp/{\rm AU})^3 \over 8 \pi^2 (M/M_\odot)} 
\left[ \omega^2 +\left( {\dot{s}\over s}\right)^2\right] 
\left( {{\rm yr} \over t_{\rm E}} \right)^2,
\label{eq11}
\end{equation}
which represents the ratio between kinetic and potential energies.
To be a bound system, the parameter should be less than unity.  
Based on the obtained orbital lensing parameters of the best-fit 
solution, we find $\eta=0.60$, implying that the result is 
consistent.

\section{Conclusion}

We analyzed the light curve of a dramatic repeating binary-lens
event OGLE-2009-BLG-092/MOA-2009-BLG-137 for which the light 
curve was characterized by two distinct peaks separated by 
$\sim 100$ days with perturbations near both peaks.  
By precisely predicting the occurrence of the second perturbation 
from the analysis of data conducted just after the first perturbation, 
we demonstrated that real-time modelings can be routinely done for
anomalous events.
Data covering the second peak obtained from follow-up observations 
enabled us to uniquely determine the physical parameters of the 
lens system.  From the analysis of the data, we found that the 
event occurred on a bulge clump giant and it was produced by a 
binary lens composed of a K and M-type main-sequence stars.  The 
estimated masses of the individual masses of the binary components 
were $M_1=0.69 \pm 0.11\ M_\odot$ and $M_2=0.36\pm 0.06\ M_\odot$, 
respectively, and they were separated in projection by $r_\perp=
10.9\pm 1.3\ {\rm AU}$.  The measured distance to the lens was 
$D_{\rm L}=5.6 \pm 0.7\ {\rm kpc}$.  
Real-time modeling of anomalous lensing events is important not 
only for efficienct use of observatinal resources but also for 
precise characterizations of lenses including planetary systems.

Besides that the lens was able to be well characterized by real-time
analysis, the event is also important because of its repeating nature. 
One of the original methods for distinguishing microlensing from 
variable stars was ``non-repeating events''.  But \citet{distefano96}
pointed out that this would cause one to miss some binary events and 
urged that selection should not be done blindly against repeating 
events. Nevertheless, there have been relatively few repeating
binary-lens events reported in the literature, e.g.\ \citet{jaroszynski08}. 
Even when they are detected, it is often difficult to distinguish them 
from those produced by binary source stars. The event 
OGLE-2009-BLG-092/MOA-2009-BLG-137 results from the special case 
where the source trajectory passes the central perturbation regions 
associated with both lens components and thus the binary nature is 
unambiguously revealed. In current lensing experiments, the majority of 
binary lenses are detected through the channel of high-magnification 
events for which the lens binarity can be easily identified from 
central perturbations \citep{han09}.  A fraction of these events will 
result in repeating events although the second peak will not be as 
dramatic as that of OGLE-2009-BLG-092/MOA-2009-BLG-137. A careful 
analysis of these repeating events can provide an independent way 
to study the statistics of wide binary stars \citep{skowron09}.

\bigskip\bigskip
We acknowledge the following support: Creative Research Initiative
Program (2009-0081561) of National Research Foundation of Korea (CH);
Korea Astronomy and Space Science Institute (C-UL);
NSF AST-0757888 (AG); NASA NNG04GL51G (BSG, AG, RWP);
Royal Society University Research Fellow (MD);
JSPS20740104 (TS); JSPS19340058 (YM);
JSPS20340052 and JSPS18253002 (MOA). 
The MiNDSTEp monitoring campaign is powered by ARTEMiS (Automated 
Terrestrial Exoplanet Microlensing Search) \citep{dominik08}.
Astronomical research at Armagh Observatory is funded by the
Department of Culture, Arts and Leisure, Northern Ireland, UK (TCH).

\end{document}